\documentclass[12pt,draft]{article}

\usepackage{amssymb}
\usepackage{latexsym}

\newtheorem{theorem}{Theorem}[section]
\newtheorem{proposition}[theorem]{Proposition}

\newtheorem{definition}[theorem]{Definition}

\newenvironment{proof}{\paragraph{\it Proof.}}{$\square$\vskip0.4cm}
\newenvironment{remark}{\paragraph{\it Remark.}}{\vskip0.4cm}

\newcommand{\nc}{\newcommand}
\nc{\HH}{{\mathbb H}}
\nc{\C}{{\mathbb C}}
\nc{\R}{{\mathbb R}}
\nc{\Z}{{\mathbb Z}}
\nc{\N}{{\mathbb N}}
\nc{\dd}{{\rm d}}
\nc{\ii}{{\bf i}}
\nc{\jj}{{\bf j}}
\nc{\kk}{{\bf k}}
\nc{\qq}{{\bf q}}
\nc{\bb}{{\bf b}}
\nc{\xx}{{\bf x}}
\nc{\ovq}{\overline{{\bf q}}}   
\nc{\ovxi}{\overline{\xi}}
\nc{\oveta}{\overline{\eta}}
\nc{\ad}{\mathop{\rm ad}\nolimits}
\nc{\tr}{\mathop{\rm tr}\nolimits}
\nc{\su}{{\mathfrak s}{\mathfrak u}(2)}
\nc{\so}{{\mathfrak s}{\mathfrak o} (4)}

\begin{document}

\title{${\rm Spin}(7)$-manifolds and symmetric Yang--Mills instantons}

\author{G\'abor Etesi
\\ {\it Yukawa Institute for Theoretical Physics,}
\\ {\it  Kyoto University,}
\\{\it Kyoto 606-8502, Japan}
\\ {\tt etesi@yukawa.kyoto-u.ac.jp}}

\maketitle

\pagestyle{myheadings}
\markright{G. Etesi: ${\rm Spin}(7)$-manifolds and symmetric instantons}

\thispagestyle{empty}

\begin{abstract}
In this Letter we establish a relationship between symmetric
$SU(2)$ Yang--Mills instantons and metrics with Spin$(7)$ holonomy. 
Our method is based on a slight extension of that of Bryant and
Salamon developed to construct explicit manifolds with special holonomies
in 1989.

More precisely, we prove that making use of symmetric $SU(2)$ Yang--Mills
instantons on Riemannian spin-manifolds, we can construct metrics on the
chiral spinor bundle whose holonomy is within Spin$(7)$. Moreover if the
resulting space is connected, simply connected and complete, the
holonomy coincides with Spin$(7)$. 

The basic explicit example is the metric constructed on the chiral spinor 
bundle of the round four-sphere by using a generic $SU(2)$-instanton
of unit action; hence it is a five-parameter deformation of the
Bryant--Salamon example, also found by Gibbons, Page and Pope. 

\end{abstract}
\centerline{Keywords: {\it {\rm Spin}$(7)$-manifolds, $SU(2)$-instantons}}
\centerline{PACS numbers: 11.10.Kk, 11.15.-q}

\section{Introduction}
The classification of holonomy groups of non-symmetric
Riemannian manifolds by M. Berger in 1955 \cite{ber}, is as fundamental
and relevant in both physics and mathematics as the classification of
simple Lie algebras by E. Cartan. 

From the mathematical point of view,
Berger's list provides a powerful and effective way to distinguish 
the main branches of Riemannian geometries. It is certainly not
an exaggeration that the main driving force of the latest decades in
Riemannian geometry is a trial for construction and understanding the
special holonomy manifolds occurring in Berger's list. The classical
example is the solution of the Calabi conjecture by Yau, which is nothing
but the proof of existence of compact Riemannian manifolds with $SU(n)$
holonomy. After solving the Calabi conjecture, the only cases had remained
in doubt were the two exceptional ones: metrics with $G_2$-holonomy in
seven dimensions and those of Spin$(7)$-holonomy in eight
dimensions. Very roughly, the
construction of these spaces took three major steps: first Bryant proved
the local existence of such metrics on open balls in $\R^7$ and $\R^8$,
respectively and also gave explicit examples in 1987 \cite{bry}. Secondly
non-compact, complete examples were found by Bryant and Salamon in
1989 \cite{bry-sal}. These spaces were re-discovered also by Gibbons,
Page and Pope in 1990 \cite{gib-pag-pop}. The next breakthrough was done
by Joyce in 1994 who constructed implicitly such metrics on plenty 
of compact manifolds and studied the moduli of theses metrics as well (for
a general and excellent introduction and outline of the topic see
\cite{joy}). 

From the physical point of view, the understanding of special holonomy
manifolds is also important. By the well-known correspondence, existence
of special-holonomy metrics on a given manifold provides us various
covariantly constant tensor fields on it, which can be interpreted as
solutions to field equations of appropriate physical theories defined over
the manifold. In simple terms, the large the symmetry of the physical
theory is, the small is the holonomy group of the underlying
manifold. Therefore, parallel to the constructions of manifolds with
more and more special holonomy by mathematicians, physicists are also
searching for such spaces for theories
with larger and larger symmetries. For example, compact Calabi--Yau spaces
are important in describing  the supersymmetric ground states of
supersymmetric ten dimensional string theories; while recently it turned
out that non-compact $G_2$-spaces are relevant in the understanding of the
unbroken $N=1$ low-energy regime of eleven dimensional M-theory
(very far from being complete cf. e.g. \cite{ach1},
\cite{ach2}, \cite{ach-wit}, \cite{ati-mal-vaf},
\cite{ati-wit}, \cite{wit}) while the less studied non-compact
Spin$(7)$-manifolds are useful tools for example in three-dimensional
$N=1$ supersymmetric Yang--Mills theory related with M-theory
\cite{guk-spa} or in brane-theory \cite{cve-gib-lu-pop}. Motivated by
this, there have been lot of efforts to construct such spaces explicitly.
Again without completeness we could mention Gibbons, Page,
Pope \cite{gib-pag-pop} and more recently a sequence of papers by Cveti\v
c et al. (as a typical example see \cite{cve-gib-lu-pop} and the
references therein) or \cite{guk-spa}. These
methods mainly are based on various coset constructions
and focus on solving the Ricci-flatness condition. However studying
other technics, e.g. based on the fundamental work \cite{bry-sal} or
more recently on \cite{bak-flo-keh} for instance, there
are indications that $SU(2)$-instantons may have an intimate relationship
with special holonomy manifolds hence would be good to find a natural
correspondence between them.  

Our paper, which is supposed to be a small step towards this direction, is
organized as follows. In Section 1 we present a slight
extension of the method of Bryant and Salamon developed in
1989 \cite{bry-sal} which allows us to construct local models for metrics
whose holonomy is within Spin$(7)$ by using ``round'' $SU(2)$ Yang--Mills
instantons on chiral spinor bundles of suitable four dimensional
Riemannian spin manifolds. Here ``round'' means that the curvatures of
these instantons are characterized by only {\it one} (i.e., not three, as
in general) smooth function. The basic example for such instantons is the
well-known five-parameter family of unit action over the round
four-sphere, hence the name. 

In Section 2 we prove via representation theory that if the resulting
space is connected, simply connected and complete, then the holonomy group
actually coincides with Spin$(7)$.

In Section 3 we turn our attention to the existence of explicit 
examples. We prove that in the case of the round four-sphere the
resulting complete examples are just deformations of the
Bryant--Salamon space \cite{bry-sal}\cite{gib-pag-pop} with moduli the
open five-ball which is the moduli space of 1-instantons. 
\vspace{0.1in}

\noindent{\bf Acknowledgement}. The work was supported by the Japan
Society for Promotion of Science, grant no. P99736.

The author thanks to Dr. T. Hausel (Department of Mathematics,
Univ. of California at Berkeley) for the useful discussions and Prof.
B.S. Acharya, S. Gukov, A. Kehagias for calling my attention to their
relevant papers in the field.

\section{Local construction of ${\rm Spin}(7)$-metrics}
Let us denote by $\HH$ the field of quaternions. In order to make our
calculations as simple as possible, we will be using quaternionic
notation: $\eta ,\xi$ etc. will denote $\HH$-valued 1-forms while $\oveta
,\ovxi$ etc. their quaternionic conjugates. Moreover we take the basic
identification $\su\cong{\rm Im}\HH$.

Let $(M,g)$ be a four dimensional Riemannian
spin-manifold. Consider a local chart $U\subset M$ and 
introduce the quaternion-valued 1-form
\[\xi :=\xi^0\pm\xi^1\ii\pm\xi^2\jj\pm\xi^3\kk\]
on it (the signs are chosen independently), where $\xi^i$ form a local
orthonormal frame on $U$ with respect to
the metric $g$. With this forms we can construct various bases for 
${\rm Im}\HH$-valued self-dual 2-forms over $(M,g)$. For example, 
the standard choice $\xi :=\xi^0 +\xi^1\ii +\xi^2\jj +\xi^3\kk$ gives rise
to the basis 
\begin{equation}
{1\over 2}\xi\wedge\ovxi =
-\left(\xi^0\wedge\xi^1+\xi^2\wedge\xi^3\right)\ii
-\left(\xi^0\wedge\xi^2-\xi^1\wedge\xi^3\right)\jj
-\left(\xi^0\wedge\xi^3+\xi^1\wedge\xi^2\right)\kk .
\label{gombbazis}
\end{equation}
Taking into account the splitting Spin$(4)\cong SU(2)\times
SU(2)$, the (complex) chiral spinor bundles $S^\pm M$ may be regarded as
$SU(2)$-bundles over $M$. Assume there is a smooth self-dual
$SU(2)$-connection i.e., an $SU(2)$-instanton $\nabla^\pm$ on $S^\pm
M$. Then $\nabla^\pm\vert_U$ can
be represented locally by ${\rm Im}\HH$-valued 1-forms $A^\pm$. Consider the
curvature $F^\pm$ of this connection, locally given by $F^\pm =\dd
A^\pm +{1\over 2}[A^\pm , A^\pm ]=\dd A^\pm +A^\pm\wedge A^\pm$. We make
the following restriction.
\begin{definition}
Let $(M, g)$ be a four dimensional Riemannian spin-manifold. We call an
$SU(2)$-instanton $\nabla^\pm$ on the chiral spinor bundle $S^\pm M$ {\rm
round}, if there is a smooth function $f^\pm : M\rightarrow\R$ and a
suitable $\HH$-valued 1-form $\xi$, constructed above, such that its
curvature can be written over all local charts as 
\begin{equation}
F^\pm ={f^\pm\over 4}\xi\wedge\ovxi .
\label{folteves}
\end{equation}
\end{definition}
The energy-density of an round instanton on $U$ has the
shape $\vert F^\pm\vert^2_g
=3(f^\pm)^2\xi^0\wedge\xi^1\wedge\xi^2\wedge\xi^3$
consequently self-duality guarantees that if $\nabla^\pm$ is not flat then
$f^\pm$ nowhere vanishes. This shows that $f^\pm$ is strictly 
positive or negative. 

Bianchi identity implies that the derivative of the round $F^\pm$
has the shape 
\begin{equation}
\dd\left({f^\pm\over 4}\xi\wedge\ovxi\right) =-A^\pm\wedge \left(
{f^\pm\over 4}\xi\wedge\ovxi\right) +{f^\pm\over 4}\xi\wedge
\ovxi\wedge A^\pm.
\label{xiderivalt}
\end{equation}
From now on we will follow the work of Bryant and Salamon
(cf. pp. 846-847 of \cite{bry-sal}) although we remark that our notations
and conventions will differ significantly from theirs. 
 
Consider the chiral spinor bundle $S^\pm M$. This is a non-compact, 
eight dimensional real manifold possessing
the structure of a two-rank complex vector bundle over $M$. We regard the 
fibers, all isomorphic to $\C^2$, as copies of $\HH$. By introducing the
linear coordinate system $(y^0, y^1, y^2, y^3)$ along each fibers, the
above identification allows us to introduce the quaternion $\qq
:=y^0+y^1\ii +y^2\jj +y^3\kk$ and consider the following $\HH$-valued
object on $S^\pm U\cong U\times\HH$: 
\[\eta :=\dd\qq +A^\pm\qq,\:\:\:\:\:\oveta =\dd\ovq -\ovq A^\pm .\]
In coordinates $\eta =\eta^0+\eta^1\ii +\eta^2\jj +\eta^3\kk$, adopted to 
(\ref{gombbazis}). We can see that under a gauge (i.e., a coordinate) 
transformation $g: M\rightarrow SU(2)\cong S^3\subset\HH$ given by
$\qq\mapsto g\qq$, $\eta$ behaves as
\[\dd\qq +A^\pm\qq\longmapsto \dd(g\qq )+(gA^\pm g^{-1}+g\dd g^{-1})g\qq
=g\dd\qq +(\dd g +gA^\pm-\dd g)\qq =g(\dd\qq +A^\pm\qq )\]
that is, it transforms as a 1-form. Therefore it is a well-defined 
$\HH$-valued 1-form over the whole chiral spinor bundle. Its derivative
is easily calculated:
\[\dd\eta =-A^\pm\wedge\eta +{f^\pm\over 4}\xi\wedge\ovxi\qq
,\:\:\:\:\:\dd\oveta =-\oveta\wedge A^\pm -{f^\pm\over 4}\ovq\xi\wedge
\ovxi .\]
In this way the derivative of the other self-dual basis
${1\over 2}\eta\wedge\oveta$ looks like 
\[\dd \left({1\over 2}\eta\wedge\oveta\right)
=-A^\pm\wedge\left( {1\over 2}\eta\wedge\oveta\right)+ {1\over
2}\eta\wedge\oveta\wedge A^\pm +{f^\pm\over 
8}\left(\xi\wedge\ovxi\wedge\qq\oveta +\eta\ovq\wedge\xi\wedge\ovxi
\right) .\]
Let us denote by $r^2:=|\qq |^2=\qq\ovq$ the radial coordinate on the
fibers; with this notation we can write $2r\dd r=\dd\qq\:\ovq
+\qq\:\dd\ovq$, implying the following identities:
\[\qq\oveta =r\dd r-r^2A^\pm,\:\:\:\:\:\eta\ovq =r\dd r+r^2A^\pm .\]
These calculations eventually yield
\begin{equation}
\dd \left({1\over 2}\eta\wedge\oveta\right) =
-A^\pm\wedge\left({1\over 2}\eta\wedge\oveta\right) +{1\over
2}\eta\wedge\oveta\wedge A^\pm+{f^\pm\over 4}r\xi\wedge\ovxi\wedge\dd r 
-{f^\pm\over 8}r^2\left( -A^\pm\wedge\xi\wedge\ovxi +\xi\wedge\ovxi\wedge
A^\pm\right) . 
\label{etaderivalt}
\end{equation}
Via the last but one equation we also prove the equality
\begin{equation}
{r^2\over 2}A^\pm\wedge\eta\wedge\oveta =r\dd r\wedge 
\eta\wedge\oveta . 
\label{rkifejezes}
\end{equation}

As a next step, we introduce the following real valued 4-forms on
$S^\pm M$, which play a crucial role in the determination of the
Spin$(7)$-structure:
\[\Omega_1:={1\over 24}{\rm Re}\left(\xi\wedge\ovxi\wedge
\overline{\xi\wedge\ovxi}\right) 
=\xi^0\wedge\xi^1\wedge\xi^2\wedge\xi^3,\]
\[\Omega_2:={1\over 4}{\rm
Re}\left(\xi\wedge\ovxi\wedge\overline{\eta\wedge\oveta}\right) =\]
\[\xi^0\wedge\xi^1\wedge\eta^0\wedge\eta^1+\xi^0\wedge\xi^1\wedge\eta^2
\wedge\eta^3 +\xi^2\wedge\xi^3\wedge\eta^0\wedge\eta^1+\xi^2\wedge\xi^3
\wedge\eta^2\wedge\eta^3\]
\[+\xi^0\wedge\xi^2\wedge\eta^0\wedge\eta^2-\xi^0\wedge\xi^2\wedge
\eta^1\wedge\eta^3-\xi^1\wedge\xi^3\wedge\eta^0\wedge\eta^2+\xi^1\wedge
\xi^3\wedge\eta^1\wedge\eta^3\]
\[+\xi^0\wedge\xi^3\wedge\eta^0\wedge\eta^3+\xi^0\wedge\xi^3\wedge\eta^1
\wedge\eta^2+\xi^1\wedge\xi^2\wedge\eta^0\wedge\eta^3+\xi^1\wedge
\xi^2\wedge\eta^1\wedge\eta^2,\]
and
\[\Omega_3:={1\over 24}{\rm 
Re}\left(\eta\wedge\oveta\wedge\overline{\eta\wedge\oveta}\right) 
=\eta^0\wedge\eta^1\wedge\eta^2\wedge\eta^3.\]
By the straightforward invariance of the definition, these forms are
well-defined on $S^\pm M$ (they are defined by the Killing-form 
$-\tr (AB)=2{\rm Re}(x\overline{y})$ on the Lie algebra
$\su\cong{\rm Im}\HH$). One has two other expressions for the
4-form $-\Omega_1+\Omega_2-\Omega_3$ (cf. p. 834 of
\cite{bry-sal}). First
we can write $-\Omega_1+\Omega_2-\Omega_3=\xi^0\wedge\zeta
+*_{\zeta}\zeta$ where
\[\zeta =\xi^1\wedge
(\xi^3\wedge\xi^2+\eta^0\wedge\eta^1-\eta^3\wedge\eta^2)+{\rm
Re}\left(
(\xi^3+\xi^2\ii )\wedge (\eta^0+\eta^1\ii )\wedge (\eta^3-\eta^2\ii
)\right) .\] 
This decomposition enables us to conclude
that the 4-form $-\Omega_1+\Omega_2-\Omega_3$ is kept fixed at one hand by
the group $\{ 1\}\times G_2\subset GL_+(8, \R )$ where the subspace
spanned by $\xi^0$ is acted on trivially while the form $\zeta$ is fixed
by the natural action of $G_2$. 
On the other hand we observe $-\Omega_1+\Omega_2-\Omega_3=-{1\over
2}\alpha\wedge\alpha +{\rm Re}\beta$ with
\[\alpha :=\xi^0\wedge\xi^1-\xi^3\wedge\xi^2-\eta^0\wedge\eta^1
+\eta^3\wedge\eta^2,\]
\begin{equation}
\beta :=(\xi^0+\xi^1\ii )\wedge
(\xi^3-\xi^2\ii )\wedge (\eta^0-\eta^1\ii )\wedge (\eta^3+\eta^2\ii ).
\label{komplex}
\end{equation}
By this representation it is also possible to see that
$-\Omega_1+\Omega_2-\Omega_3$ remains invariant under the group
$SU(4)\subset GL_+(8, \R )$ where 
the complex structure on the tangent spaces is induced by 
the complex 4-form $\beta$. These observations yield that the full
stabilizer of $-\Omega_1+\Omega_2-\Omega_3$ is the
group Spin$(7)\subset GL_+(8, \R )$, as it is proved in \cite{bry-sal} or
in a more detailed way in \cite{bry}. 

First note that taking into account (\ref{xiderivalt}), 
(\ref{etaderivalt}) and (\ref{rkifejezes}) we have (cf. p. 847 of
\cite{bry-sal})
\begin{equation}
\dd\Omega_1=0,\:\:\:\:\:\dd\Omega_3 ={r\over 2}f^\pm\Omega_2\wedge\dd r. 
\label{azonossagokI}
\end{equation}
Moreover by writing $\Omega_2=(f^\pm )^{-1}{\rm
Re}\left( F^\pm\wedge\overline{\eta\wedge\oveta}\right)$, one obtains
\[\dd\Omega_2= -{\dd f^\pm\over (f^\pm)^2}\wedge{\rm
Re}\left( F^\pm\wedge\overline{\eta\wedge\oveta}\right)+{1\over
f^\pm}\dd{\rm
Re}\left( F^\pm\wedge\overline{\eta\wedge\oveta}\right) .\]
But in light of (\ref{xiderivalt}), (\ref{etaderivalt}) we can
write $\dd {\rm Re}\left( F^\pm\wedge\overline{\eta\wedge\oveta}\right)=
3r(f^\pm )^2\Omega_1\wedge\dd r$
leading to 
\begin{equation}
\dd\Omega_2=-{\dd f^\pm\over f^\pm}\wedge\Omega_2 +
3rf^\pm\Omega_1\wedge\dd r.
\label{azonossagokII}
\end{equation}
(cf. p. 847 of \cite{bry-sal}). Moreover we have the two straightforward
equalities
\begin{equation}
\Omega_1\wedge\dd f^\pm =0,\:\:\:\:\:\Omega_3\wedge\dd r=0.
\label{azonossagokIII}
\end{equation}
\begin{remark}
We can always assume that $f^\pm$ is positive
in (\ref{azonossagokI}) and (\ref{azonossagokII}) because the
transformations $\xi\wedge\ovxi\mapsto
-\xi\wedge\ovxi$ and $\eta\wedge\oveta\mapsto -\eta\wedge\oveta$ leave
$\Omega_i$ invariant while one has $f^\pm\mapsto -f^\pm$. 
\end{remark} 
Now consider two functions $\varphi ,\psi : S^\pm M\rightarrow\R^+$ and
assume that they depend on the fiber coordinates $y^i$ only through
the radial coordinate $r$. Take the 4-form
\[\Omega :=-\varphi^2\Omega_1+\varphi\psi\Omega_2-\psi^2\Omega_3\]
and the associated metric $g_\Omega$, locally given by
\[\dd s^2:=\varphi\left(\xi_0^2+\xi_1^2+\xi_2^2+\xi_3^2\right)
+\psi\left(\eta_0^2 +\eta_1^2 +\eta_2^2 +\eta_3^2\right) .\]
$\Omega$ is self-dual with respect to the associated metric, moreover at
each tangent spaces we can find an isomorphism sending 
$\Omega$ into $-\Omega_1+\Omega_2-\Omega_3$. As it is proved for example
in \cite{bry},\cite{bry-sal} or \cite{joy}, if $\nabla\Omega =0$ with
respect to $g_\Omega$ (a non-linear problem!) then this metric has a
holonomy group, whose identity component ${\rm Hol}^0(g_\Omega )$ is
contained within Spin$(7)$. Now we prove that by a suitable
choice of the functions $\varphi$ and $\psi$ we can achieve this. 
\begin{proposition}
Let $(M,g)$ be a Riemannian spin four-manifold and $\nabla^\pm$ an
round $SU(2)$-instanton on the spinor bundle
$S^\pm M$. Then there is a metric $g_\Omega$ on the non-compact
eight-manifold $S^\pm M$, locally given by
\begin{equation} 
\dd s^2= (1+r^2)^{3/5}f^\pm\left(\xi_0^2+\xi_1^2+\xi_2^2+\xi_3^2\right)+
{4\over 5}(1+r^2)^{-2/5}\left(\eta_0^2 +\eta_1^2 +\eta_2^2
+\eta_3^2\right) ,
\label{spin7metrika}
\end{equation}
where $f^\pm$ comes from {\rm (\ref{folteves})}, satisfying ${\rm
Hol}^0(g_\Omega )\subseteq{\rm Spin}(7)$. The space $(S^\pm M, g_\Omega
)$ is complete if $(M,g)$ is compact. If $(M,g)$ is non-compact but
complete and
\begin{equation}
\int\limits_0^\infty \sqrt{f^\pm (\gamma (t))}\:\dd t=\infty
\label{teljesseg}
\end{equation}
for each curve $\gamma :\R^+\rightarrow M$ (not contained in any compact
set of $M$) then $(S^\pm M, g_\Omega )$ is also complete.  
\end{proposition}

\begin{proof} By virtue of Theorem 2.3 of \cite{bry-sal}, it is enough to
prove that with a suitable choice of the functions $\varphi$, $\psi$, the
4-form $\Omega$ is closed i.e. $\dd\Omega =0$. 

Let us denote by $(x^0, x^1, x^2, x^3)$ a local coordinate
system on $U\subset M$. Calculating the exterior derivative we get
\[\dd\Omega =-2\varphi\left({\partial\varphi\over\partial
r}\dd
r+{\partial\varphi\over\partial x^i}\dd x^i\right)
\wedge\Omega_1-\varphi^2\dd\Omega_1\]
\[+\left(\varphi{\partial\psi\over\partial r}\dd r
+\varphi{\partial\psi\over\partial x^i}\dd x^i
+\psi{\partial\varphi\over\partial r}\dd r +
\psi{\partial\varphi\over\partial x^i}\dd
x^i\right)\wedge\Omega_2+\varphi\psi\:\dd\Omega_2\]
\[-2\psi\left({\partial\psi\over\partial r}\dd
r+{\partial\psi\over\partial x^i}\dd
x^i\right)\wedge\Omega_3-\psi^2\dd\Omega_3.\]
By using identities (\ref{azonossagokI}), (\ref{azonossagokII}) and
(\ref{azonossagokIII}) this reduces to
\[\dd\Omega =-\left({\partial\varphi^2\over\partial
r}-3rf^\pm\varphi\psi\right)\dd
r\wedge\Omega_1+\left({\partial (\varphi\psi )\over \partial r}-{r\over
2}f^\pm\psi^2\right)\dd r\wedge\Omega_2\]
\[-{\partial \varphi^2\over\partial x^i}\dd x^i\wedge\Omega_1+{\partial
(\varphi\psi )\over\partial x^i}\dd
x^i\wedge\Omega_2-\varphi\psi{\dd f^\pm\over
f^\pm}\wedge\Omega_2-{\partial\psi^2\over\partial x^i}\dd
x^i\wedge\Omega_3.\]
The terms of the first row are eliminated by solving the system of
ordinary differential equations just appeared in the coefficients. As we
have seen we can assume $f^\pm >0$ hence the general solution is
(cf. p. 847
of \cite{bry-sal})
\[\varphi (r, x^i) ={1\over
h_1(x^i)}(h_1(x^i)r^2+h_2(x^i))^{3/5}f^\pm (x^i),
\:\:\:\:\:\psi (r, x^i)={4\over 5}(h_1(x^i)r^2+h_2(x^i))^{-2/5}\]
where $h_1$, $h_2$ are arbitrary functions of $x^i$'s only.

Now focus on the second row of the above expression. Take simply
$h_1=h_2=1$. In this case $\psi$ is independent of $x^i$ consequently the
last term of the second row vanishes. Moreover by noticing that with the
above choice of $h_i$ we have
\[{\partial\varphi^2\over\partial x^i}\dd
x^i\wedge\Omega_1=2(1+r^2)^{6/5}f^\pm\dd f^\pm\wedge\Omega_1,\]
we can see that by the first equation of
(\ref{azonossagokIII}) the first term vanishes, too. Henceforth the
calculation amounts to an expression for
the middle terms (after substituting $\varphi$, $\psi$):
\[\dd\Omega ={4\over 5}(1+r^2)^{1/5}\dd f^\pm\wedge\Omega_2-{4\over
5}(1+r^2)^{1/5}\dd f^\pm\wedge\Omega_2 =0\]
showing $\dd\Omega =0$ with the above choice of the functions $\varphi$
and $\psi$. This implies that the associated metric satisfies ${\rm
Hol}^0(g_\Omega )\subseteq{\rm Spin}(7)$.

Now we turn our attention to the geodesic completeness of the resulting
metric (\ref{spin7metrika}). We can see that this metric is geodesically
complete along each fibers because $\eta_0^2 +\eta_1^2+\eta_2^2+\eta_3^2$
is complete and 
\[\int\limits_0^\infty {\dd r\over (1+r^2)^{1/5}}=\infty .\]
Consequently (\ref{spin7metrika}) is complete if
$f^\pm\left(\xi_0^2+\xi_1^2+\xi_2^2+\xi_3^2\right)$ is complete; this is
valid if $M$ is compact. However it might fail this property if $M$ is not
compact and $f^\pm$ decays too fast. Suppose
$M$ is non-compact and $\xi_0^2+\xi_1^2+\xi_2^2+\xi_3^2$ is complete on
it; then the re-scaled metric is complete if and only if
(\ref{teljesseg}) is valid. Consequently $(S^\pm M, g_\Omega )$ might be 
incomplete.
\end{proof}
In summary we have found a local form (\ref{spin7metrika}) of Riemannian
metrics $g_\Omega $ with the property ${\rm Hol}^0(g_\Omega )\subseteq
{\rm Spin}(7)$. 

\section{Proof of ${\rm Spin}(7)$ holonomy}

We have to still find a condition for the holonomy groups
actually coincide with ${\rm Spin}(7)$. By using a result of Bryant and
Salamon, we can prove this.
\begin{proposition}
Assume that $M$ is connected and simply connected and the associated
space $(S^\pm M, g_\Omega)$ of the previous proposition is complete. Then
${\rm Hol}(g_\Omega )\cong{\rm Spin}(7)$ is valid.
\end{proposition}
\begin{proof} If $M$ is connected and simply connected then the
same is true for $S^\pm M$. Therefore, by the aid of Theorem
2.4 of Bryant and Salamon \cite{bry-sal} we have to show that there are no
non-trivial parallel 1-forms and 2-forms on $(S^\pm M, g_\Omega )$ because
this implies that the holonomy ${\rm Hol}^0(g_\Omega )={\rm Hol}(g_\Omega
)$ cannot be smaller than Spin$(7)$. 

To achieve this, we list all the possible holonomy groups which are
subgroups of Spin$(7)$ (cf. Theorem 10.5.7 in \cite{joy}):
\begin{enumerate}
\item[(i)] Reducible actions:
\[\begin{array}{ll}
               \{ 1\} & \mbox{acting on $\R^8$ trivially,} \\
               \{ 1\}\times SU(2)\cong\{ 1\}\times{\rm Spin}(3) &
\mbox{acting on $\R^8\cong\R^4\oplus\C^2$ trivially on $\R^4$, as
usual on $\C^2$}, \\
               SU(2)\times SU(2)\cong{\rm Spin}(4) & \mbox{acting on
$\R^8\cong\C^2\oplus\C^2$ on each $\C^2$ as usual,} \\
               \{ 1\}\times SU(3) & \mbox{acting on
$\R^8\cong\R^2\oplus\C^3$ trivially on $\R^2$, as usual on $\C^3$,} \\
               \{ 1\}\times G_2   & \mbox{acting on
$\R^8\cong\R\oplus\R^7$ trivially on $\R$, as usual on $\R^7$.}
                        \end{array}\]

\item[(ii)] Irreducible actions:
\[\begin{array}{ll} 
             Sp(2)\cong{\rm Spin}(5) & \mbox{acting on $\R^8\cong\HH^2$
as usual,} \\
             SU(4)\cong{\rm Spin}(6) & \mbox{acting on $\R^8\cong\C^4$ as
usual,} \\
             {\rm Spin}(7) & \mbox{acting on $\R^8$ as usual.}
\end{array}\]
\end{enumerate}
Assume now $(S^\pm M, g_\Omega )$ is moreover complete. Then
(\ref{spin7metrika}) is a complete metric on a simply
connected manifold and does not split, as
one can check. Taking into account the de Rham theorem \cite{deR}, its
holonomy group cannot act reducibly on the tangent spaces. Consequently
the groups listed in (i) cannot occur.
             
Concerning part (ii) of the list, we can proceed as follows: assume the
holonomy group is Spin$(6)$ only. The group
Spin$(6)\cong SU(4)$ acts irreducibly on $\R^8\cong\C^4$ i.e. there are
no non-trivial parallel 1-forms on the space $(S^\pm M, g_\Omega
)$ (cf. e.g. Theorem 2.5.2 of \cite{joy}). Moreover we have an
induced action on $\Lambda^2\C^4$ as well, which
gives rise to an action on $\Lambda^2\R^8$. Since this action is nothing
but one of the fundamental representations of $SU(4)$, it is also
irreducible. Consequently, there are no non-trivial parallel 2-forms,
too. But this implies that the holonomy group must be Spin$(7)$, a
contradiction.

Now assume that the holonomy group is Spin$(5)$. The action of
Spin$(5)\cong Sp(2)$ is also irreducible on $\R^8\cong\HH^2$ i.e. again
there are no parallel 1-forms. The induced action
on $\Lambda^2\HH^2$ is {\it not} irreducible, however. To see this,
we will follow \cite{bro-tom}, pp. 269-272. Consider the identification 
$\HH^2\cong\C^4$ with a basis $(e_{\pm 1}, e_{\pm 2})$, and regard the
action of $Sp(2)$ as a subgroup of $SU(4)$, leaving the well-known   
symplectic form invariant. With this notation, the induced
reducible representation of $Sp(2)$ on $\Lambda^2\C^4$ splits as
$\Lambda^2\C^4\cong V_1\oplus V_5$
where the action is trivial on the first summand $V_1\cong\C$, spanned by
the 2-form
\[e^*_1\wedge e^*_{-1}+e^*_2\wedge e^*_{-2}\]
($e^*_{\pm i}$ are the dual basis elements to $e_{\pm i}$), while the five
dimensional orthogonal complement $V_5$ (with respect to
the standard Hermitian inner product on $\C^4$) is acted on
non-trivially. This induces a splitting of $\Lambda^2\R^8$, too. Therefore
we can see that a non-trivial parallel 2-form on $(S^\pm M, g_\Omega )$
must be either the real or imaginary part of the 2-form 
\[f((\xi^0+\xi^1\ii
)\wedge (\xi^3-\xi^2\ii )+(\eta^0-\eta^1\ii )\wedge
(\eta^3+\eta^2\ii ))\]
where the identification $\R^8\cong\C^4$ on the
tangent spaces is induced by (\ref{komplex}); $f$ is some complex valued
function on $S^\pm M$. But we can check that
the only 2-form of the above shape which is parallel with respect to
(\ref{spin7metrika}) is the zero  2-form. Indeed, since $\nabla
(\xi^i\wedge\xi^j)\not= 0$ and depends only on $x^i$
furthermore $\nabla (\eta^i\wedge\eta^j)\not= 0$ and depends on both $x^i$ 
and $\qq$, this implies that $f$ must be zero.

Hence again we have not been able to find non-trivial parallel 1- and
2-forms consequently the holonomy must be Spin$(7)$.
\end{proof}

\section{A global example: the round four-sphere}

In this section we construct new explicit examples whose holonomy
groups are Spin$(7)$. 

The most straightforward example is the round four-sphere
$(S^4, g)$ with isometry group $SO(5)$ (\cite{fre-uhl},
pp. 99-105). Because of conformal invariance,
we may consider the flat $\R^4\cong\HH$ as well. Let $\xx
,\bb\in\HH$ and $\lambda >0$ real. Then the basic instanton together
with its curvature looks like
\[A = {\rm Im}{\xx\dd\overline{\xx }\over 1+|\xx
|^2},\:\:\:\:\:F={\dd\xx\wedge\dd\overline{\xx}\over
(1+\vert\xx\vert^2)^2}.\] 
If we apply the homothety $T_{\lambda
,\bb}: \xx\mapsto\lambda (\xx
-\bb )$ then we get the five-parameter family of instantons,
\[T^*_{\lambda ,\bb}A:=A_{\lambda , \bb }={\rm Im}{(\xx
-\bb)\dd\overline\xx \over 
\lambda^2+\vert\xx -\bb\vert^2}, \:\:\:\:\:T^*_{\lambda ,\bb}F:=F_{\lambda
,\bb}= {\lambda^2\dd\xx\wedge\dd\overline{\xx}\over
(\lambda^2+\vert\xx -\bb\vert^2)^2}.\]
Therefore these instantons are round with respect to the self-dual basis
(\ref{gombbazis}). Now putting $A_{\lambda , \bb }$ into
(\ref{spin7metrika}) we can produce a five-parameter family of complete
metrics $g_\Omega$ over $S^\pm S^4$ with holonomy within Spin$(7)$: 
\[\dd s^2= {\lambda^2(1+r^2)^{3/5}\over (\lambda^2+\vert\xx
-\bb\vert^2)^2}(\dd x_0^2+\dd x_1^2+\dd x_2^2+\dd x_3^2)+
{4\over 5}(1+r^2)^{-2/5}\left(\eta_0^2 +\eta_1^2 +\eta_2^2
+\eta_3^2\right) =\]
\[ {\lambda^2(1+r^2)^{3/5}(1+|\xx |^2)^2\over (\lambda^2+\vert\xx
-\bb\vert^2)^2}\dd\Omega^2_{S^4}+{4\over 5}(1+r^2)^{-2/5}\left(\eta_0^2
+\eta_1^2 +\eta_2^2 +\eta_3^2\right) \]
where we have used the conformal re-scaling $1/(1+|\xx |^2)^2(\dd
x_0^2+\dd x_1^2+\dd x_2^2+\dd x_3^2)=\dd\Omega^2_{S^4}$. Taking the
inverse of the homothety $T_{\lambda ,\bb}$ which sends $A_{\lambda ,\bb}$
back to $A$, we just recover the Bryant--Salamon metric on the chiral
spinor bundle of the four-sphere \cite{bry-sal}, also found by Gibbons,
Page and Pope \cite{gib-pag-pop}:
\[\dd s^2=(1+r^2)^{3/5}\dd\Omega^2_{S^4}+{4\over
5}(1+r^2)^{-2/5}\left(\eta_0^2
+\eta_1^2 +\eta_2^2 +\eta_3^2\right) .\] 
This procedure intuitively corresponds to
the limit $\lambda\rightarrow\infty$ i.e., the basic instanton 
$A$ is ``centerless''. Consequently these new
spaces are deformations of the Bryant--Salamon space with moduli space the
five-ball $B^5$, nothing but the moduli space of $SU(2)$-instantons of
unit charge on $S^4$. In this picture the Bryant--Salamon space
corresponds to the centerless instanton represented by the center of
$B^5$. By the previous proposition, these spaces have holonomy Spin$(7)$.

\section{Concluding remarks}
A very natural question arises whether it is possible to remove the
very restrictive "roundness" assumption for the Yang--Mills instantons in
this extended Bryant--Salamon construction. If yes, we could establish a
correspondence between $SU(2)$ Yang--Mills instantons over compact spin
manifolds and Spin$(7)$-metrics on the chiral spinor bundle. If the
underlying spin manifold is non-compact then the geodesic completeness of
the associated space would depend on the fall-off properties of the field
strength of the instanton.

Of course it would be also interesting to know if the above method can be
repeated for the $G_2$-case. The main difference between the two cases is
that while for Spin$(7)$ we have only one non-linear partial differential
equation for the existence, in the $G_2$ case we have two; consequently
it is typically more difficult to obey the conditions for the
$G_2$-case.

\end{document}